\documentclass[11pt]{article}

\usepackage{amsmath,amsfonts,amssymb}
\usepackage{graphicx}
\usepackage{secdot}

\author{\footnotesize Shahabeddin Mostafanazhad Aslmarand$^1$, Warner A. Miller$^1$, Paul M. Alsing$^2$\and \footnotesize Verinder S. Rana${}^3$}
\date{\footnotesize %
   \itshape{ $^1$Department of Physics, Florida Atlantic University, Boca Raton, FL, 33431, USA}\\%
    $^2$Air Force Research Laboratory, Information Dire
   ctorate, Rome, NY, 13441, USA\\%
    $^3$SPAWAR Systems Center Pacific 53560 Hull St Building 605, San Diego, CA 92152, USA\\%
     \today
}

\begin{document}
\title{Quantum Reactivity: A Measure of Quantum Correlation\thanks{Work supported by AFOSR and AFOSR/AOARD Grant \#FA2386-17-1-4070 }
}

\maketitle

\begin{abstract}
Defining a robust measure of quantum correlation for multipartite states is an unresolved and challenging problem.  Existing measures of quantum correlation are either not scalable or do not satisfy all the accepted properties of a measure of quantum correlation.  We introduce a novel geometric measure of quantum correlation that we refer to as quantum reactivity.  This measure is extendable  to an arbitrary large number of qubits and  satisfies the required properties of monotonicity and invariance under unitary operations.  Our approach is based on generalization of Schumacher's singlet state triangle inequality that used an information geometry--based entropic distance.  We define quantum reactivity as the familiar ratio of surface area to volume. To accomplish this, we use a generalization of information distance to area, volume and higher--dimensional volumes.   We examine a spectrum of multipartite states (Werner, W, GHZ etc.) and demonstrate that the quantum reactivity measure is a monotonic function for quantum correlation which satisfies all the properties of a measure for quantum correlation, and provides an ordering of these quantum states as to their degree of correlation.
\end{abstract}


 \newpage                    
\section{Measures for Entanglement and Correlations}
\label{Sec:Intro}

Quantum correlation is widely recognized as a key resource for quantum computing, and quantifying this resource is important for this field \cite{Preskill:2012}.    Quantum correlation encompasses a broader class of correlations than quantum entanglement.  While it is important to continue to explore entanglement measures as a primary focus as the resource for quantum computing.  Nevertheless, states that are not entangled can contain quantum correlations, and this stronger than classical correlation may be advantageous for some  quantum computing applications, e.g. quantum non--locality without entanglement \cite{Bennett:1999,Ali:2010,Horodecki:2005,Nisent:2006}.  In addition it was demonstrated that separable states can potentially scale more favorably than classically allowed \cite{Lanyon:2008}.   Therefore, in this manuscript we will concentrate on the broader class of measures of quantum correlation.   Among many applications, the utility of quantum correlation is clearly revealed in quantum cryptography \cite{Ekert}, quantum dense coding \cite{Charles H. Bennett}, and quantum teleportation \cite{Charles H Bennett}.  Furthermore, quantum correlation has the potential to optimize classical communication by reducing its complexity \cite{Cleve}.  For example,  the computational power of a quantum network is related to its  degree of quantum correlation \cite{Bennett:1999}.  These examples suggest that a scalable measure of correlation that can identify regions containing a high degree of quantum correlation within a quantum network can be important for quantum computing.  However, this has proven to be a formidable problem.  In particular, quantum correlation in multipartite states has exponentially increasing complex structure --- the correlations exist in a Hilbert space of exponentially--large dimension \cite{Vedral}.

It is widely accepted that such a measure of quantum correlation should satisfy following two properties \cite{Horedecki1}.
\begin{enumerate}
\item It should be monotonic, i. e. non-increasing under local operations and classical communication {\fontfamily{lmss}\selectfont (LOCC)}.
\item It should be invariant under the unitary operators.
\end{enumerate}

\begin{figure}[h!]
 \centering
\vspace{20pt}
  \includegraphics[width=5 in]{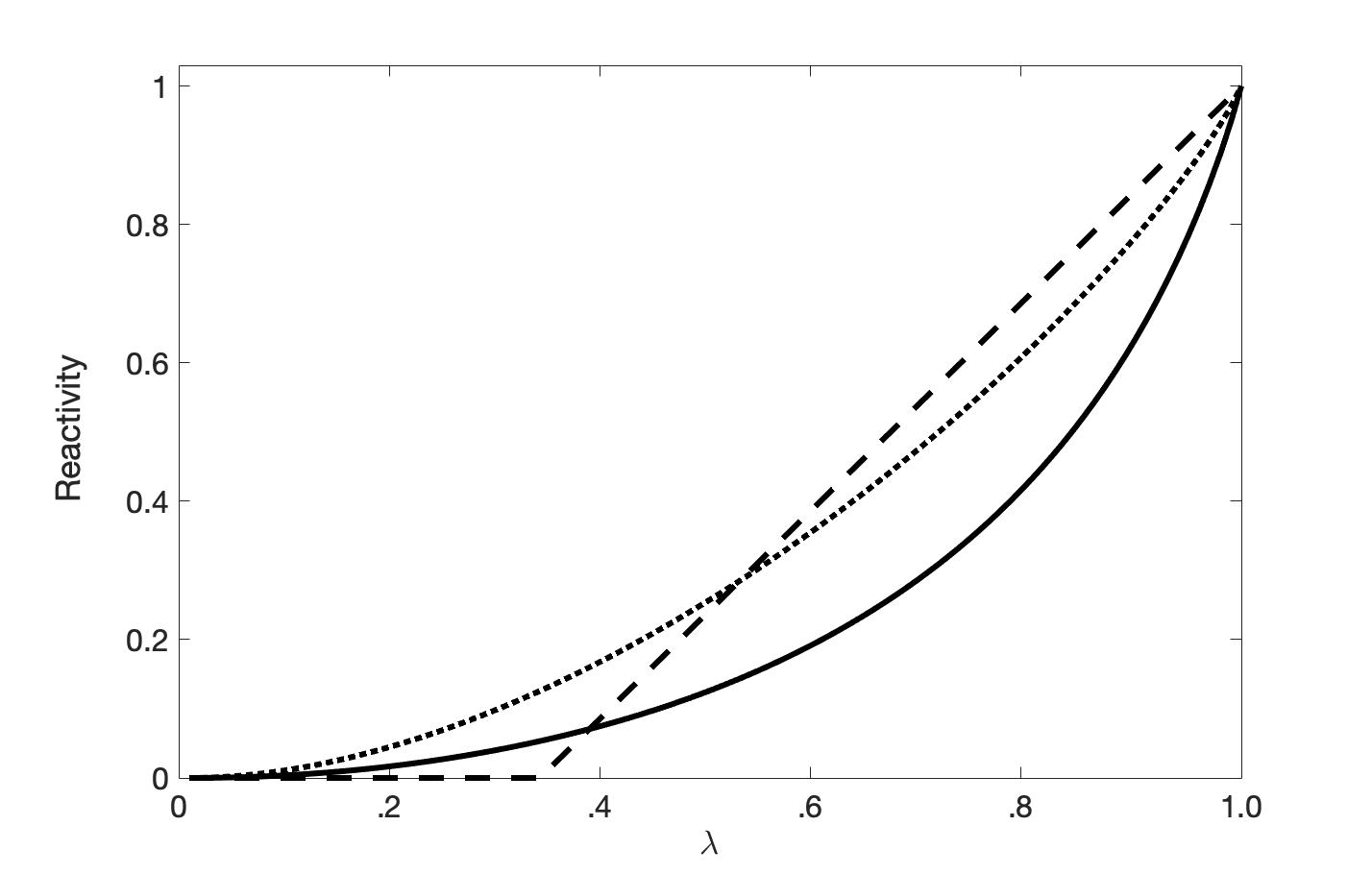}
\hspace{+15pt}
 \vspace{0pt}
 \caption{ This figure shows the change in the reactivity(solid line), discord(dotted line), concurrence(dash line) with respect to the entanglement parameter $\lambda$.  State is maximally entangled when $\lambda=1$. This will be discussed  in detail in Sec.~\ref{sec:werner}. }
  \label{Fig:Werner}
 \end{figure}

It's possible to partition measures of correlation into two different class, measures that capture quantum correlation directly, and measures that are capturing quantum entanglement. We briefly summarize these two classes of measures by focusing on concurrence and discord,  and we also suggest that the quantum reactivity measure proposed Sec.~\ref{sec:reactivity} might be a useful tool to probe quantum correlation.  In Sec.~\ref{sec:reactivity} we also provide a comparison of the three measures (concurrence, discord and reactivity) for a bipartite Werner state as illustrated in Fig.~\ref{Fig:Werner}  
 
Concurrence is perhaps the most well known in the class of measures of entanglement, and it captures precisely the onset of entanglement.  This measure of entanglement was first suggested by Scott Hill and William K. Wootters \cite{Wottward}.  Concurrence is an entanglement monotone defined for a mixed state of two qubits, where 
\begin{equation}
C(\rho ):= \max{\left({0,\sqrt{\lambda_1}-\sqrt{\lambda_2} - \sqrt{\lambda_3}-\sqrt{\lambda_4}}\right)},
\end{equation}
and $ \left\{\lambda {_j}\right\}_{j=1,2,3,4}$ are the corresponding eigenvalue of $\widetilde{\rho}$ in descending order ($\lambda_1\ge \lambda_2\ge \lambda_3\ge\lambda_4$), and  
$\widetilde{\rho}  :=  ( {\sigma_ y} {\otimes} \sigma ){\rho }^*  (\sigma _ y \otimes \sigma_ y)$.\\
The motivation behind this definition of concurrence is to provide a measure of the degree of 
non--separability of the quantum state as a distance. The concurrence is zero when the state is separable and when it is non zero it provides a measure of how far this state is from achieving separability.  

The second measure we outline is quantum discord.  Discord was introduced by Ollivier and Zurek and is an entropic measure \cite{Zurek}.  It captures quantum correlation even when there is no quantum entanglement.  They illustrated this on the Werner state (Fig.~\ref{Fig:Werner}).  Their definition was motivated by the incompatibility of classical and quantum mutual information.  In particular, the two equivalent classical expressions for mutual information,
\begin{equation}
\label{Eq:I}
I (A; B) := H (A) + H (B) - H (A,B),
\end{equation}
and
\begin{equation}
\label{Eq:J}
J (A; B) = H (A) - H (A|B),
\end{equation}
differ when applied to quantum states.  Here $H(A)$ is the information entropy, $H(A, B)$ the joint entropy and $H(A|B)$ the conditional entropy. The conditional entropy measures the uncertainty left in A after the value of B is known (presumably by a measurement).  Eqs.~\ref{Eq:I} and \ref{Eq:J} are equal to each other for classical probability distributions; however, when we calculate them for quantum systems using von Neumann entropy 
\begin{equation}
I(\rho_{AB}) = S (\rho_A) + S (\rho_B) - S (\rho_{AB})
\end{equation}
\begin{equation}
J_{A}(\rho_{AB} )=S(\rho _{B})-S(\rho_{AB|\Pi^A})
\end{equation}
$J$ and $I$ are not ordinarily equal.  This disparity led to the definition of quantum  discord 
\begin{equation}
 {D}_{A}(\rho ):L=I(\rho )-\max _{\{\Pi _{j}^{A}\}}J_{\{\Pi _{j}^{A}\}}(\rho ).
 \end{equation}
The most noticeable characteristic of discord is that it can be non zero for a separable state. Thus, the  absence of entanglement is not necessarily equivalent to the absence of quantum correlation.

They are many measures of quantum correlations in addition to discord,  and there are many measures for entanglement other than concurrence.\cite{Horedecki1,Alber:2001,Rulli}  However,  these measures each have their own problems. Concurrence measures entanglement but is not sensitive to the quantum correlations that may exist in separable states,  and it's not scalable to a larger number of qubits.  Discord, on the other hand, is scalable to a larger number of qubits but it's not necessarily non-increasing under LOCC.  Consequently, we feel it is important to explore different approaches to this problem, especially given its  importance to the field of quantum computing. Therefore, we introduce quantum reactivity in this manuscript. 

In this paper we focus on a unique geometric rendering of entanglement that was first  introduced by Schumacher \cite{Schumacher}.  He constructed a geometry based on the measurement outcomes on an ensemble of  identical bipartite quantum states.  The quadrilateral geometry he constructed captured the entanglement of the singlet state through a Shannon-based entropic measure of distance.  This may seem surprising at first; however, he introduced a pair of detectors for each of the two entangled qubits.  Alice measures one of the two qubits by choosing randomly between two detectors labeled $A_1$ and $A_2$, and Bob does the same with detectors labeled $B_1$ and $B_2$. This enabled him to construct the quadrilateral with edges, $\overline{A_1B_1}$, $\overline{B_1A_2}$, $\overline{A_2B_2}$ and $\overline{A_1B_2}$, where Schumacher demonstrated that for certain settings of the detectors, the direct distance between $A1$ and $B_2$ is longer than the sum of the three other indirect distances.  Entanglement is related to the geometry!  

We generalize Schumacher's bipartite geometry to multipartite systems by extending the definition of  information distance, too information area, volume and higher--dimensional volumes \cite{Miller:2018,Miller:2019Venn}. This allows one to calculate the quantum reactivity that we define as the average of surface area divided by average over volume. The average is taken over over all possible measurement basis.  We show here that  the quantum reactivity is sensitive to quantum correlations, and it is expressed using Shannon-based entropic volumes defined over the space of measurement outcomes.  In an accompanying manuscript \cite{Aslmarand:2019spie} , we showed that the quantum reactivity is (1) a monotonically increasing function of quantum correlation, i. e. a non-increasing under LOCC, (2)  invariant under unitary transformations.  The  mean averaging makes our measure independent of choice of measurement basis, in other words,  our measure of correlation only depend on the inherent correlation of quantum state.  In this manuscript we introduce our geometric measure of quantum reactivity.  In Sec.~\ref{sec:EEG}, we describe a generalization of Schumacher's geometry to higher-dimensional multipartite quantum states. Following this, in Sec.~\ref{sec:example}, we provide an illustrative example of a generalization of the Schumacher geometry for the the $|GHZ\rangle$  tripartite state.  In Sec.~\ref{sec:reactivity} we define reactivity for a multipartite state, and we apply this to a spectrum of multipartite states (bipartite and tripartite Werner state, four-qubit Werner state, as well as a modified Werner state).  Finally, in Sec.~\ref{sec:fini} we discuss future questions that we will explore by using quantum reactivity.  

 \section{Quantum Geometry in the Space of Measurements}
 \label{sec:EEG}
 
\begin{figure}[h!]
\centering
\includegraphics[width=4 in]{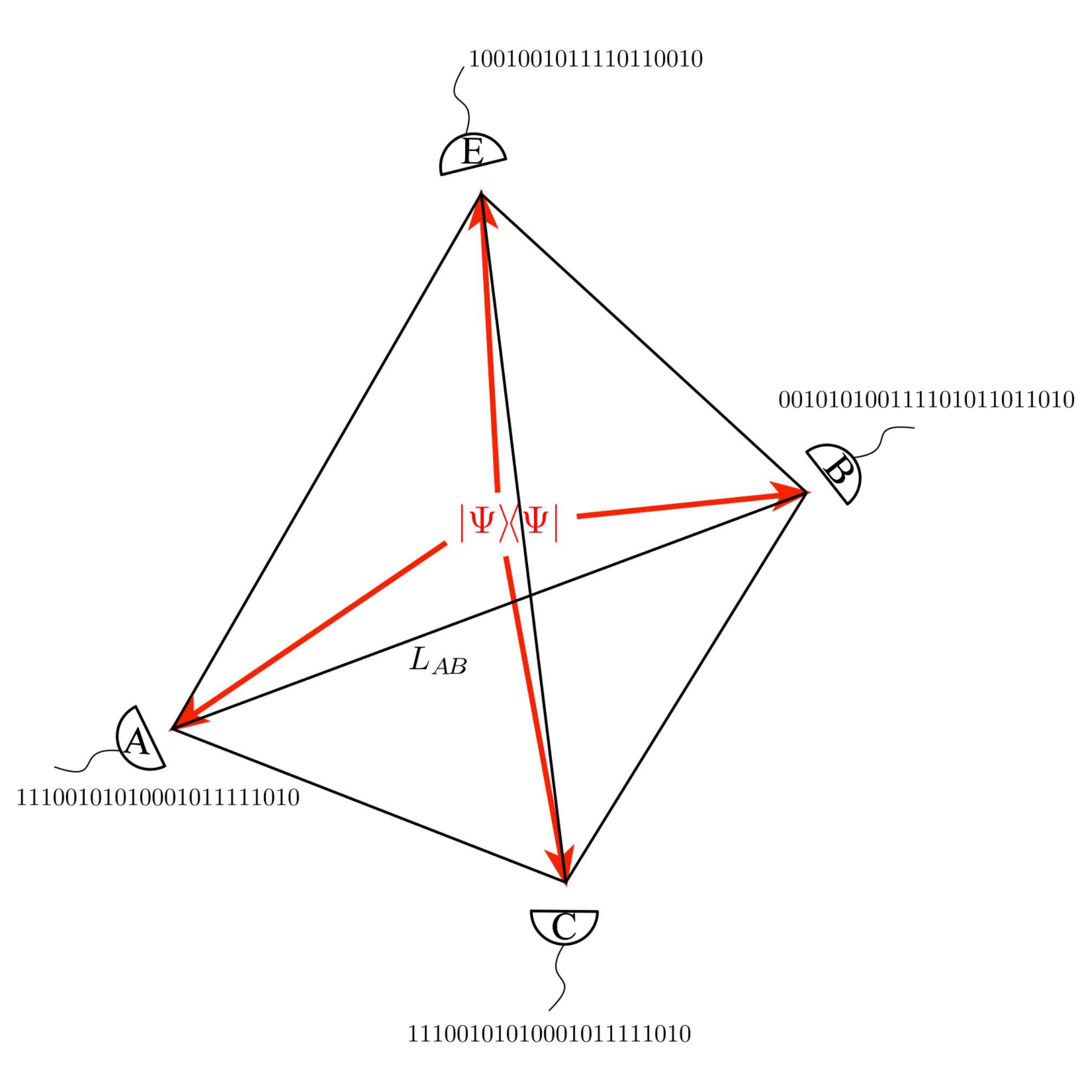}
\caption{An illustration of a simplicial geometry representation of a four--qubit state illustrated by the four photons emanating from the center of the tetrahedron (arrows).  The tetrahedral  geometry emerges from a recording of measurements made by $A$, $B$, $C$ and $E$ over an ensemble of identically prepared states.  At each observer their string of $1$'s and $0$'s form a binary random variable, the combination forms a joint probability distribution from which we can define the complete geometry of the tetrahedron. Here tetrahedron is  parameterized by the eight Stoke parameters of the detectors of our four observers.}
\label{Fig:SOM}
 \end{figure}

 The assertion of John Archibald Wheeler  that ``No elementary quantum phenomenon is a phenomenon until it is brought to close by an irreversible act of amplification.''  was inspired by  Niels Bohr.  This point of view, together with the Principle of Complementarity,  is at the very heart of Wheeler's {\em It-from-Bit} framework \cite{Wheeler:1990}.  Here we explore a spectrum of quantum networks within this information-centric geometric landscape. The quantum network is a multipartite state with $d$ qubits that can contain entanglement and be mixed.  Observers create a ``space of measured data'' from repeated experiments over a set of identically-prepared quantum states.  Each observer records a ``$1$'' if their detector triggers, otherwise a ``$0$" is recorded.  This generates a string of $1$'s and $0$'s at each detector as illustrated in Fig.~\ref{Fig:SOM}.  The string of numbers can be represented by a binary random variable.  The observers may have more than one detector, and therefore each observer may acquire more than one binary random variable. Once these random variables are formed, we can apply an information geometry measure of distance, area, volume and $d$-volumes to the network of observers \cite{Miller:2019Venn,Rolkin,Rajski,Miller,Zurek:1989}. Following Schumacher, these measures are defined using the familiar Shannon expression for conditional entropy that will be described in this section \cite{Schumacher,Shannon}. 
 
 What is unique to our definition of quantum information geometry is that we project the qudit state into a classical space of  a $d$-dimensional joint probability distribution. We make repeated measurements on an ensemble of identically-prepared quantum states to produce this probability distribution. From this distribution we can calculate all other marginal or joint probability distributions. In this sense we are constructing a functional from the quantum density matrix to a classical distribution; we are essentially ``{\em probing the quantum state within a space of measurements.}''\cite{Miller:2018} We describe this in detail below after we discuss novel entropic geometry constructions. 

We construct our reactivity based on information geometry \cite{Miller:2019Venn}.  The fundamental quantity of any information theory is information entropy.  The entropy, $H(X)$ is a function of the probability spectrum of values of a random variable $X$ . Similar to Schumacher, we use Shanon's entropy \cite{Shannon} where 
\begin{equation}
\label{Eq:SE}
H(X)=- \sum_{i=1}^s { p(x_i) \ln{p(x_i)} },
\end{equation}
for an $s$--state random variable.  Here $p(x_i) = p(X\!=\!x_i)$ is the probability that the random variable $X$ has the value $x_i$.  Probability measures uncertainty about the occurrence of a single event, but entropy provides a measure of the uncertainty of a collection of events.  In other words,  the $H(X)$ is a measure of the uncertainty associated with the probability distribution over $X$.

The entropy is the largest when our uncertainty of the value of the random variable is complete (e.g. uniform distribution of probabilities), and the entropy is zero if the random variable always takes on the same value. These bound the possible values for entropy, 
\begin{equation}
0{\leq}H(X_i){\leq}log(s).
\end{equation}
For example, consider an electron with two possible configurations, up and down.  In the most general quantum state is the normalized  superposition state
\begin{equation}
|\psi\rangle = c_1 |\!\uparrow\rangle + c_2 |\!\downarrow \rangle
\end{equation}  
\begin{equation}
P(x_i) = \left\{  
\begin{array}{lll}
\uparrow & \hbox{with probability} &{\mid c_1\mid }^2\\
\downarrow & \hbox{with probability} & {\mid {c_2} \mid }^2
\end{array}
\right..
\end{equation}
The entropy of this 2--state probability  distribution is 
\begin{equation}
 {H} (X)=-\sum _{i=1}^{2}{\mathrm {P} (x_{i})\log _{2}\mathrm {P} (x_{i})}=- {\mid c_1\mid }^2\log _{2}({\mid c_1\mid }^2)-{\mid c_2\mid }^2\log _{2}({\mid c_2\mid }^2).
 \end{equation}
 In the case that ${\mid {c_2} \mid }^2 = {\mid {c_1} \mid }^2 =\frac{1}{2}$,  $H(X)$ will equal to 1.
This entropy obeys a set of special  properties for an $s$--state random variable:
\begin{enumerate}
\item $ H(X) {\geq} 0$, with equality if one outcome has probability 1.
\item $H(X){\leq} \ln s$, where $s$ is the number of possible outcomes
for X, with equality if each outcome has probabilities
$1/s$.
\item $ H(X,Y) {\geq} H(X)$, where $H(X,Y)$ is the joint information
for $X$ and $Y$.
\item $ H(X,Y){\leq }H(X)+H(Y)$ with equality if $X$ and $Y$ are independent.
\end{enumerate}
These properties make it possible to define an information geometry measure.  It is also useful to use the  conditional entropy
\begin{equation}
\label{eq:ce}
H(X|Y) = H(X,Y)-H(Y)
\end{equation} 
that measures the uncertainty of $X$ after $Y$ is known. We  use these entropies to build our geometrical structure.  There are many possible definitions;  however, in this paper we follow Schumacher, and use an extension of the Shannon-based information distance defined by Rokhlin\cite{Rolkin} and Rajsk\cite{Rajski} (RR) where
 \begin{equation}
 \label{Eq:D}
 {\mathcal D}_{AB} = H_{A|B} + H_{B|A} = 2H_{AB}-H_{A}-H_{B}.
 \end{equation}
 This information distance is a proper metric since,
\begin{enumerate}
\item It is constructed to be symmetric, ${\mathcal D}_{AB} = D_{BA}$
\item It obeys the triangle inequality, ${\mathcal D}_{AB}\geq {\mathcal D}_{AC}+ {\mathcal D}_{CB}$.
\item It is non-negative, ${\mathcal D}_{AB}\geq 0$, and equal to 0 when $A$“=”$B$.
\end{enumerate}
In the case that $A$ and $B$ are uncorrelated  
\begin{equation}
H_{AB}=H_{A}+H_{B} 
\end{equation}
and the distance obtains its maximal permissible value
\begin{equation}
 {\mathcal D}_{AB} =H_{A}+H_{B} .
 \end{equation}
 Our choice of metric is not unique.  Different choices can give distinct behavior \cite{Thiago}.  In this manuscript we use the RR metric.  

In addition to the information distance, we can analogously assign an information area to geometry of measurement space,\cite{Miller:2018,Miller:2019Venn,Miller}  where 
\begin{equation}
\label{Eq:A}
{\mathcal A}_{ABC} = H_{A|BC}H_{B|CA}+H_{B|CA}H_{C|AB}+H_{C|AB}H_{A|BC}.
\end{equation}
This can be generalized to higher-dimensional simplexes, e.g.  the information volume for a tetrahedron 
\begin{equation}
\label{Eq:V}
\begin{array}{ll}
{\mathcal V}_{ABCD}& := H_{A|BCD}H_{B|CDA}H_{C|DAB} +H_{B|CDA}H_{C|DAB}H_{D|ABC}\\
&\ \ +H_{C|DAB}H_{D|ABC}H_{A|CDB}+H_{D|ABC}H_{A|BCD}H_{B|CDA}
\end{array}
\end{equation}
is a natural generalization of Eq.~\ref{Eq:A}.\cite{Miller:2018,Miller:2019Venn,Miller}.

For classical probability distributions, we showed that these formulas are well defined and have all of the requisite symmetries, positivity, bounds and structure usually required for such formula \cite{Miller:2019Venn}. Each of these geometric measures are bounded,
\begin{equation}
0\leq D_{AB} \leq H_A+H_B \leq 2 \log s,
\end{equation}  
\begin{equation}
0\leq {\mathcal A}_{ABC} \leq 3 (\log s)^2,
\end{equation}  
\begin{equation}
0\leq {\mathcal V}_{ABCD} \leq 4 (\log s)^3.
\end{equation}  
The minimum values occur when the random variables are completely correlated, and their maximum values obtained when the random variables are completely uncorrelated.

We use the information distance, area and volume defined on the ``space of measurements''  to define a curvature--based definition of quantum correlation that we refer to as reactivity.   We have found that such curvature measures provide a relationship between the curvature of the ``space of measurements'' and the quantum correlation of the underlying quantum system.  We find that quantum entanglement is reactive.

 \section{An Illustrative Example:  The Information Geometry of a Three Qubit State}
 \label{sec:example}

We find it useful to introduce our approach by focusing on the maximally--entangled qutrit state, namely: 
\begin{equation}
\label{Eq:state}
|GHZ\rangle = \frac{1}{\sqrt{2}} \left( |111 \rangle +  |000\rangle\right).
\end{equation}
For this state we can calculate the information geometry using Eqs.~\ref{Eq:D}--\ref{Eq:V}.  This is a three qubit state. One qubit is sent to Alice ($A$), one to Bob ($B$) and the third to Charlie ($C$).  $A$, $B$  and $C$ prepare a large ensemble ($N$ copies) of identically prepared states Eq.~\ref{Eq:state}, The three observers each choose a measurement basis. They then measure the initial state (Eq.~\ref{Eq:state}) with their choice of detector settings.  The observers record their digital measurements.  In particular they record a  "1" if their detector registers an event,  otherwise they record a "0" and keep their order in time.  This procedure produces an ordered string of $0$'s and $1$'s of  length $N$.  After  these measurements, $A$, $B$ and $C$ each has generated a binary random variable.  Together, $A$, $B$ and $C$ form a joint probability distribution, 
\begin{equation}
\label{eq:jp}
p(A=a_i,B=b_i,C=c_i)\ \ \hbox{with} \ a_i,b_i,c_i\in\{0,1\}.
\end{equation}
This gives the probability that $A$'s detector, $B$'s detector and $C$'s detector will measure the values $a_i$, $b_i$ and $c_i$; respectively.  From this joint distribution we can define entropy and extract the information geometry.  We can calculate the three information distances as well as the information area formed by $A$, $B$ and $C$.  

Let us  suppose $A$ oriented her detector at the Stokes angles $\theta_a$ and $\varphi_a$, that $B$ oriented his apparatus at angles $\theta_b$ and $\varphi_b$ and $C$ oriented his apparatus at angels $\theta_c$ and $\varphi_c$.  The measurement operator for each of these detectors are
 \begin{equation}
 M_A =  \frac{1}{2 } \left (I+\cos{(\theta_a)} \sigma_z +\sin(\theta_a)\sin(\varphi_a) \sigma_y+\sin{(\theta_a)} \cos(\varphi_a) \sigma_x\right) \otimes I  \otimes I,
 \end{equation}
\begin{equation}
 M_B = I\otimes \frac{1}{2 } \left (I+\cos{(\theta_b)} \sigma_z +\sin(\theta_b)\sin(\varphi_b) \sigma_y+\sin{(\theta_b)} \cos(\varphi_b) \sigma_x\right)  \otimes I,
\end{equation}
\begin{equation}
 M_C = I\otimes I \otimes \frac{1}{2 } \left (I+\cos{(\theta_c)} \sigma_z +\sin(\theta_c)\sin(\varphi_c) \sigma_y+\cos{(\varphi_c)}\sin{(\theta_c)} \sigma_x\right).
\end{equation}

These measurements directly yield the joint probability (Eq.~\ref{eq:jp}), as well as all the pairwise probabilities, e.g. $p(A=a_i,B=b_i)$ and individual probabilities, e.g. $p(A=a_i)$.  This in turn is sufficient to determine all the entropies and conditional entropies.  Without loss of generalization, we choose to  simplify our expressions by fixing $A$'s detector to the Stokes parameters $\theta_a=\varphi_a=0$. The eight joint probabilities from the three measurements on this entangled state $M_A\otimes M_B\otimes M_C|GHZ\rangle$ are:
\begin{eqnarray}
\label{eq:jointGHZ}
p(A=1,B=1,C=1) &=& \frac{1}{2} \cos^2(\theta_b)\cos^2(\theta_c),\\
p(A=0,B=1,C=1) &= &\frac{1}{2} \sin^2(\theta_b)\sin^2(\theta_c), \\
p(A=1,B=1,C=0) &=&  \frac{1}{2}\cos^2(\theta_b)\sin^2(\theta_c),\\
p(A=0,B=1,C=0) &= & \frac{1}{2}\sin^2(\theta_b)\cos^2(\theta_c), \\
p(A=1,B=0,C=1) &= & \frac{1}{2}\sin^2(\theta_b)\cos^2(\theta_c),\\
p(A=0,B=0,C=1) &= & \frac{1}{2}\cos^2(\theta_b)\sin^2(\theta_c), \\
p(A=1,B=0,C=0) &= & \frac{1}{2}\sin^2(\theta_b)\sin^2(\theta_c),\\ 
p(A=0,B=0,C=0) &= & \frac{1}{2}\cos^2(\theta_b)\cos^2(\theta_c)
\end{eqnarray}
Tracing these joint probability over each observer yields the twelve pairwise joint probabilities,
\begin{eqnarray}
p(A=1,B=1) &=& \frac{1}{2}\cos^2(\theta_b),\\
p(A=1,B=0) &= &\frac{1}{2}\sin^2(\theta_b), \\
p(A=1,C=1) &= &\frac{1}{2}\cos^2(\theta_c),\\
p(A=1,C=0) &= &\frac{1}{2}\sin^2(\theta_c),\\
p(A=0,B=1) &= &\frac{1}{2}\sin^2(\theta_b),\\
p(A=0,B=0) &= &\frac{1}{2}\cos^2(\theta_b), \\
p(A=0,C=1) &= &\frac{1}{2}\sin^2(\theta_c),\\
p(A=0,C=0) &= &\frac{1}{2}\cos^2(\theta_c),\\
p(B=1,C=1) &= &\frac{1}{2}\left( \cos^2(\theta_b)\cos^2(\theta_c) + \sin^2(\theta_b)\sin^2(\theta_c) \right), \\
p(B=1,C=0) &= &\frac{1}{2}\left( \cos^2(\theta_b)\sin^2(\theta_c) + \sin^2(\theta_b)\cos^2(\theta_c) \right), \\
p(B=0,C=1) &= & \frac{1}{2}\left( \sin^2(\theta_b)\cos^2(\theta_c) + \cos^2(\theta_b)\sin^2(\theta_c) \right), \\
p(B=0,C=0) &= & \frac{1}{2}\left( \cos^2(\theta_b)\cos^2(\theta_c) + \sin^2(\theta_b)\sin^2(\theta_c) \right). 
\end{eqnarray}
Finally, tracing the joint probability over all pairs of observers gives us the six probabilities for the measurement outcomes of $A$, $B$ and $C$ to be 
\begin{equation}
p(A=a_i) =p(B=b_i)=p(C=c_i)=\frac{1}{2}.
\end{equation}
From these probabilities we find the entropies.   The joint entropy $H_{ABC}$ of our three observers is 
\begin{eqnarray}
H_{ABC} = 1&-&\cos^2(\theta_b)\cos^2(\theta_c) \log( \cos^2(\theta_b)\cos^2(\theta_c))\nonumber\\
&-&\cos^2(\theta_b)\sin^2(\theta_c) \log(\cos^2(\theta_b)\sin^2(\theta_c))\nonumber \\
&-& \sin^2(\theta_b)\cos^2(\theta_c) \log(\sin^2(\theta_b)\cos^2(\theta_c))\nonumber\\
&-& \sin^2(\theta_b)\sin^2(\theta_c) \log(\sin^2(\theta_b)\sin^2(\theta_c)).
\end{eqnarray}

The three joint entropy between each pair of our observers are,
\begin{eqnarray}
\label{eq:pe1GHZ}
H_{AB} &= &1-\sin^2(\theta_b) \log(\sin^2(\theta_b)) -\cos^2(\theta_b) \log(\cos^2(\theta_b)),\\
H_{AC} &= &1-\sin^2(\theta_c) \log(\sin^2(\theta_c)) -\cos^2(\theta_c) \log(\cos^2(\theta_c)),
\end{eqnarray}
\begin{eqnarray}
H_{BC} &= &1-\left( \cos^2(\theta_b)\cos^2(\theta_c) + \sin^2(\theta_b)\sin^2(\theta_c) \right)\nonumber\\ 
&&\log \left( \cos^2(\theta_b)\cos^2(\theta_c) + \sin^2(\theta_b)\sin^2(\theta_c) \right)\nonumber\\
&&\ \ \ - \left( \sin^2(\theta_b)\cos^2(\theta_c) + \cos^2(\theta_b)\sin^2(\theta_c) \right)\nonumber\\
&& \log (\left( \sin^2(\theta_b)\cos^2(\theta_c) + \cos^2(\theta_b)\sin^2(\theta_c) \right).
\end{eqnarray}
and finally the three entropies of our observers are equal
\begin{equation}
H_A=H_B=H_C=1
\end{equation}

The lengths of the three edges of the triangle formed by $A$, $B$ and $C$ for this $|GHZ\rangle$ state can be calculated using these  entropies and Eq.~\ref{Eq:D}. Similarly, the entropic area is defined using  Eq.~\ref{Eq:A} and the relation for conditional entropy in Eq.~\ref{eq:ce}.

We use the same procedure when we calculate the information geometry for the two--qubit and three--qubit Werner state in the next section, as well as for the four--qubit Werner state.  The procedure is the same for higher dimensional multipartite states. 

 \section{Reactivity as a measure for quantum correlation}
 \label{sec:reactivity}

In Sec.~\ref{Sec:Intro} we made reference to numerous measures of quantum correlation for a two--qubit states, and we highlighted two of them, discord and  concurrence.  In this section, we define our geometric  measure of correlation that we call quantum reactivity.  We also show that reactivity yields results that are qualitatively similar to the other measures.  We demonstrate this for a spectrum of known states.  However, unlike discord and concurrence, our measure allows is straightforwardly scalable to a higher number of qubits.  

Reactivity of a chemical agent is naturally defined as the ratio of its surface area to volume, and is related to its curvature. We adopt the same definition, and apply it to the information area and volumes outlined  in Sec.~\ref{sec:EEG} through Eqs.~\ref{Eq:A}-\ref{Eq:V} and their extensions to higher dimensions in \cite{Miller:2019Venn}.  In particular, a  natural definition of curvature--based reactivity can be defined for an $d$--qubit simplectic  geometry as
\begin{equation}
\label{eq:dreactivity}
{\mathcal R} := \frac{ 
\langle {}^{{}^{(d-2)}}\!{\mathcal A} \rangle_{{}_{\mathcal M}}
}{ 
\langle {}^{{}^{(d-1)}}\!{\mathcal V} \rangle_{{}_{\mathcal M}}
}.
\end{equation}
The brackets, $\langle {}^{{}^{(d)}}\!{\mathcal V} \rangle_{{}_{\mathcal M}}$,  denote the  volume--averaged mean of the information volume, ${}^{{}^{(d)}}\!{\mathcal V}$,  over all possible measurements $\mathcal M$.  In particular, 
\begin{equation}
\langle {}^{{}^{(d)}}\!{\mathcal V} \rangle_{{}_{\mathcal M}} := \frac{1}{V_{\!{}_{\mathcal M}}} \left(\int_{\mathcal M} {}^{{}^{(d)}}\!{\mathcal V}\right),
\end{equation}
and  $V_{\mathcal M} =\langle 1\rangle_{{}_{\mathcal M}} =  \left( 2 \pi^2 \right)^d$ is the volume of the ``space of measurements.'' Here we have, without loss of generalization, fixed one of the detectors at a fixed point on its Block sphere. We integrate over all measurement settings on the remaining ($d-1$) qubits. 

Our proposed geometrical measure of correlation satisfies the following properties:
\begin{enumerate}
\item This geometrical measure is monotonically increasing as quantum correlation increases.
\item This geometrical measure is invariant under unitary LOCC's.
\item This geometrical measure is non-increasing under LOCC's.
\item The maximum bound on this curvature can't be obtained using only classical correlation.
\end{enumerate}
These properties have been reported in an accompanying manuscript.\cite{Aslmarand:2019spie}

In the next three subsections, we examine the reactivity for the following states: (1) the mixed bipartite Werner state; (2) the mixed three--qubit Werner state and a less--correlated modified--Werner state; and finally (3) the  four-qubit generalization of the Werner state. 

\subsection{Reactivity of a bipartite Werner State}
\label{sec:werner}

For illustrative purposes, we first consider the reactivity of the bipartite Werner state.  This state has a controllable parameter, $\lambda$, that governs the degree of entanglement and quantum correlation,
\begin{equation}
\label{Eq:Werner}
\rho_{{}_{Werner}}=\lambda \,|\psi_{singlet}\rangle \langle\psi_{singlet} |+ {\frac{1-\lambda}{4}}\, I.
\end{equation}
This state is separable when $\lambda\le 1/3$, totally mixed when $\lambda=0$ and the maximally entangled singlet state when $\lambda=1$.\cite{Werner}   

We imagine that an observer Alice ($A$) makes measurements on one to the qubits, and BoB ($B$) makes measurements on the other qubit.  Without loss of generalization we can permanently fix  the Stokes parameters of $A$ to zero, i.e. $\theta_a=\varphi_a=0$.  However, $B$ makes measurements over his Bloch sphere with each measurement characterized by its two Stokes angles $\theta_b$ and $\varphi_b$. The reactivity is defined on the emergent geometry of the space of measurements.  While reactivity may be more intuitive when applied to a four--qubit state; nevertheless, we can apply this to a two--qubit state equally well by using Eq.~\ref{eq:dreactivity} with $d=2$,  where we find 
\begin{equation}
{\mathcal R}:={\frac{1}{\langle D_{AB}\rangle_{{}_{\mathcal M}}}}.
\end{equation}
Here the weighted average of the distance is defined over the 2--dimensional Bloch--sphere measurements, 
\begin{equation}
\langle D_{AB}\rangle_{{}_{\mathcal M}}= \frac{1}{V_{\!{}_{\mathcal M}}}  \left( 
\int_{\mathcal M}  {\mathcal D}_{AB} \right)
:=  \frac{1}{2\pi^2} \int_0^{\pi}  \int_0^{2\pi}  
{\mathcal D}_{AB}(\theta_b,\varphi_b)\, d\theta_b d\varphi_b.
\end{equation}
and yields the reactivity defined over all possible relative combinations of detector orientations ($\mathcal M$) for $A$ and $B$.  

Each value $\lambda$ of the entanglement parameter in Eq.~\ref{Eq:Werner} yields a unique value for the reactivity, ${\mathcal R}$.  The function ${\mathcal R}(\lambda)$ with respect to correlation parameter $\lambda$ is a monotonically increasing function as illustrated in Fig.~\ref{Fig:Werner}.  Its minimum is the totally mixed state $\rho_{\!{}_{Werner}}(\lambda=0)=\frac{1}{4}I$,  and it obtains its max value at $\lambda=1$  when the state is the maximally entangled singlet state, $|\psi_{singlet}\rangle$.  As our state gets more correlated as $\lambda$ increases the geometry of the space of measurements is more reactive (or curved).  

The behavior of the reactivity measure shares the essential characteristics of  Zurek \& Ollivier's discord measure as illustrated in Fig.~\ref{Fig:Werner}.  We plotted a linear transformation of the reactivity to make ${\mathcal R}(0)=0$ (totally mixed)and ${\mathcal R}(1)=1$ (maximally entangled bipartite state).  Both discord and reactivity yield non-zero quantum correlations even below $\lambda=1/3$ when concurrence shows us that there is no entanglement. Quantum correlation can still exist even in separable states that are not entangled can contain quantum correlations, and this stronger than classical correlation may be advantageous for some  quantum computing applications, e.g. quantum non--locality without entanglement \cite{Bennett:1999,Ali:2010,Horodecki:2005,Nisent:2006}.  In addition it was demonstrated that separable states can potentially scale more favorably than classically allowed \cite{Lanyon:2008}. We emphasize that even though the behavior of discord and reactivity are qualitatively similar, that the two measures of quantum correlation are fundamentally different, and this difference may be useful.  
  
\subsection{Reactivity of the tripartite Werner State and modified Werner State }
\label{sec:werner3}
 
In this section, we explore the behavior of reactivity for two different  tripartite states, the Werner state and a modified Werner state. Both states are mixed states. The density matrix for the three--qubit Werner state
\begin{equation}
\label{eq:ws}
\rho_1=\lambda \,|GHZ\rangle \langle GHZ |+ {\frac{1-\lambda}{8}}\, I
\end{equation}
is constructed as a weighted sum of the density matrix for the maximally entangled $|GHZ\rangle$ state,
\begin{equation}
|GHZ\rangle = \frac{1}{\sqrt{2}} \left(
|000\rangle+|111\rangle
\right),
\end{equation}
and the totally mixed state $I$. When $\lambda=0$ the state is the separable mixed state with least correlation and when $\lambda=1$ it is the maximally entangled state. The parameter $\lambda$ is the entanglement parameter. 
We compare the reactivity for this Werner state and a modified Werner state. The density matrix for the modified Werner state is based on the less--entangled $|W\rangle$ state, where
 \begin{equation}
\label{eq:mws}
\rho_2=\lambda \,|W\rangle \langle  W |+ {\frac{1-\lambda}{8}}\, I.
\end{equation} 
Here too, this density matrix is constructed as a weighted sum of the density matrix for the less--entangled  $|W\rangle$ state,
\begin{equation}
|W\rangle = \frac{1}{\sqrt{3}} \left(
|001\rangle+|010\rangle + |100\rangle 
\right),
\end{equation}
and the uncorrelated  mixed state $I$. When $\lambda=0$ the state is the maximally mixed state and when $\lambda=1$ it is the highly--entangled $|W\rangle$ state. 

The three--qubit reactivity is equally- well defined defined over the ``space of measurements'' on an ensemble of identically--prepared quantum states made by three observers, $A$, $B$ and $C$.   The reactivity is defined by Eq.~\ref{eq:dreactivity} by setting $d=3$.  The three observers form a triangle.  As they record their measurements from an ensemble of identically--prepared quantum states they will generate a joint probability distribution 
\begin{equation}
p\left(A =a_i,B=b_i,C=c_i\right),
\end{equation} 
from whence we can calculate all the entropies ($H_A$), joint entropies ($H_{AB}$ and $H_{ABC}$) and conditional entropies ($H_{A|B}$ and $H_{A|BC}$) as we did in Sec.~\ref{sec:example}.  These entropies define the information distance and  area using Eqs.~\ref{Eq:D} and  \ref{Eq:A}; respectively. Then the reactivity for this qutrit system is defined to be the ratio of the weighted average of the  perimeter of the triangle given by Eq.~\ref{Eq:D}
\begin{equation}
\langle {\mathcal P}\rangle_{{}_{\mathcal M}}= 
\langle {\mathcal D}_{AB}+{\mathcal D}_{AC}+{\mathcal D}_{BC} \rangle_{{}_{\mathcal M}}
\end{equation}
to the weighted average of the area of the triangle given by Eq.~\ref{Eq:A} 
\begin{equation}
\langle {\mathcal A}\rangle_{\!{}_{\mathcal M}}:= \frac{1}{V_{\!{}_{\mathcal M}}} \left( \int_{\mathcal M} {\mathcal A}\right),
\end{equation}
where 
\begin{equation}
\label{eq:3reactivity}
{\mathcal R} := \frac{ 
\langle {\mathcal P}\rangle_{{}_{\mathcal M}}
}{
\langle {\mathcal A}\rangle_{\!{}_{\mathcal M}}
}.
\end{equation}
Here the averages are taken over over all detector settings $\mathcal M$ with $V_{\!{}_{\mathcal M}}=(2\pi^2)^2$ for the area and $V_{\!{}_{\mathcal M}}=(2\pi^2)$ for the perimeter.  Without loss of generalization we can again set $A$'s detector at $\theta_a=\varphi_a=0$ for all measurements. The integration that defines the reactivity will then be a function of the four Stokes parameters for $B$ and $C$, namely $\theta_b$, $\varphi_b$ and $\theta_c$, $\varphi_c$; respectively.
\begin{figure}[h]
 \centering
\hspace*{0cm}
  \includegraphics[width=5 in]{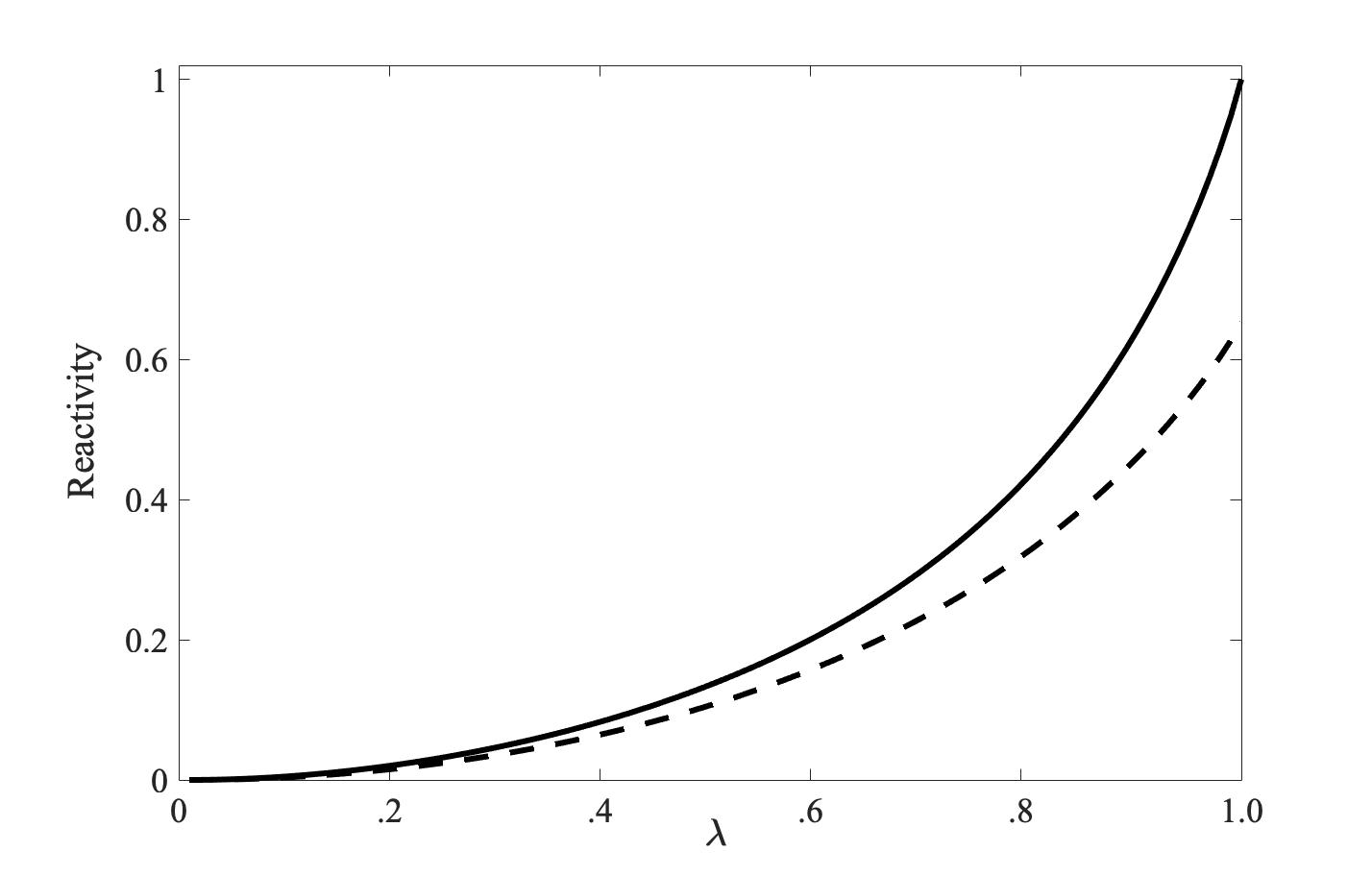}
 \caption{The reactivity, ${\mathcal R}$  is plotted as a function of the entanglement parameter $\lambda$ for the three--qubit Werner state(solid line) defined by Eq.~\ref{eq:ws},  and the less correlated modified Werner State(dashed line) defined in Eq.~\ref{eq:mws}.  The modified Werner state is formed using the  $|W\rangle$ state instead of  maximally--entangled qutrit $|GHZ\rangle$ state.}
 \label{Fig:WW}
 \end{figure}
 Fig.~\ref{Fig:WW} illustrates the relationship of the reactivity for the Werner state and the modified Werner state.  The curve for the modified Werner state in Fig.~\ref{Fig:WW} lies below the Werner state.  This is intuitive as there is more entanglement in the $|GHZ\rangle$ state than in the $|W\rangle$ state. 
 
\subsection{Reactivity for the Four Qubit Werner State}
\label{sec:ghz4}

Reactivity  is scalable in the number of qubits in the sense that it is an analytic function of the volume and its bounding surface. It is straight--forwardly extendable to higher dimensions.  We demonstrate the scaling to higher number of qubits by calculating the reactivity  for a four--qubit  state.  In particular, we examine tin the section the four--qubit Werner state 
\begin{equation}
\label{Eq:WWWW}
\rho_{Werner}=\lambda \,|GHZ_4\rangle \langle GHZ_4 |+ {\frac{1-\lambda}{16}}\, I,  
\end{equation}
where
\begin{equation}
| GHZ_4\rangle =  \frac{1}{\sqrt{2}} \left( |0000\rangle + |1111\rangle \right).
\end{equation}
For this state we will have four observers $A$, $B$, $C$ and $E$ measuring this state. They form a tetrahedron as illustrated in Fig.~\ref{Fig:SOM}.  Once again, and without loss of generality, we can permanently align $A$'s detector along the $z$-axis and calculate the joint probability distributions $p_{ABCE}$ from the measurements on an ensemble of identically prepared states made by $B$, $C$ and $E$.  From this distribution we can define all the entropies as well as the information volume and the surface area for this tetrahedral geometry. This calculation mirrors the calculation we did in  Sec.~\ref{sec:werner3} except it is in one higher dimension.  The reactivity for this four--qubit system is defined to be the ratio of the weighted average of the tetrahedrons surface to the weighted average of the tetrahedral volume.  The weighted average of the surface is the sum of the weighted average of the four bounding triangle areas
\begin{equation}
\langle {\mathcal S}\rangle_{{}_{\mathcal M}}= 
\langle {\mathcal A}_{ABC}+{\mathcal A}_{ABE}+{\mathcal A}_{ACE}+{\mathcal A}_{BCE}  \rangle_{{}_{\mathcal M}}.
\end{equation}
where each of the weighted triangle areas is given bu  Eq.~\ref{Eq:A}.
The weighted average of the tetrahedral volume is given by Eq.~\ref{Eq:V}  and gives 
\begin{equation}
\langle {\mathcal V}\rangle_{\!{}_{\mathcal M}}:= \frac{1}{V_{\!{}_{\mathcal M}}} \left( \int_{\mathcal M} {\mathcal A}\right),
\end{equation}
Where $V_{\!{}_{\mathcal M}}=(2\pi^2)^3$.
We obtain the reactivity for this qutrit state 
\begin{equation}
\label{eq:4reactivity}
{\mathcal R} := \frac{ 
\langle {\mathcal S}\rangle_{{}_{\mathcal M}}
}{
\langle {\mathcal V}\rangle_{\!{}_{\mathcal M}}
}.
\end{equation}
using Eq.~\ref{eq:dreactivity} with $d=4$. Here the averages are taken over over all detector settings relative to $A$ that forms the ``space of measurements'' $\mathcal M$.  The integration that defines the reactivity will then be a function of the six Stokes parameters for $B$, $C$ and $E$, namely $\theta_b$, $\varphi_b$, $\theta_c$, $\varphi_c$ and $\theta_e$, $\varphi_e$; respectively.
 
\begin{figure}[h]
 \centering
 \vspace{20pt}
\hspace{50 pt}
  \includegraphics [width=4 in]{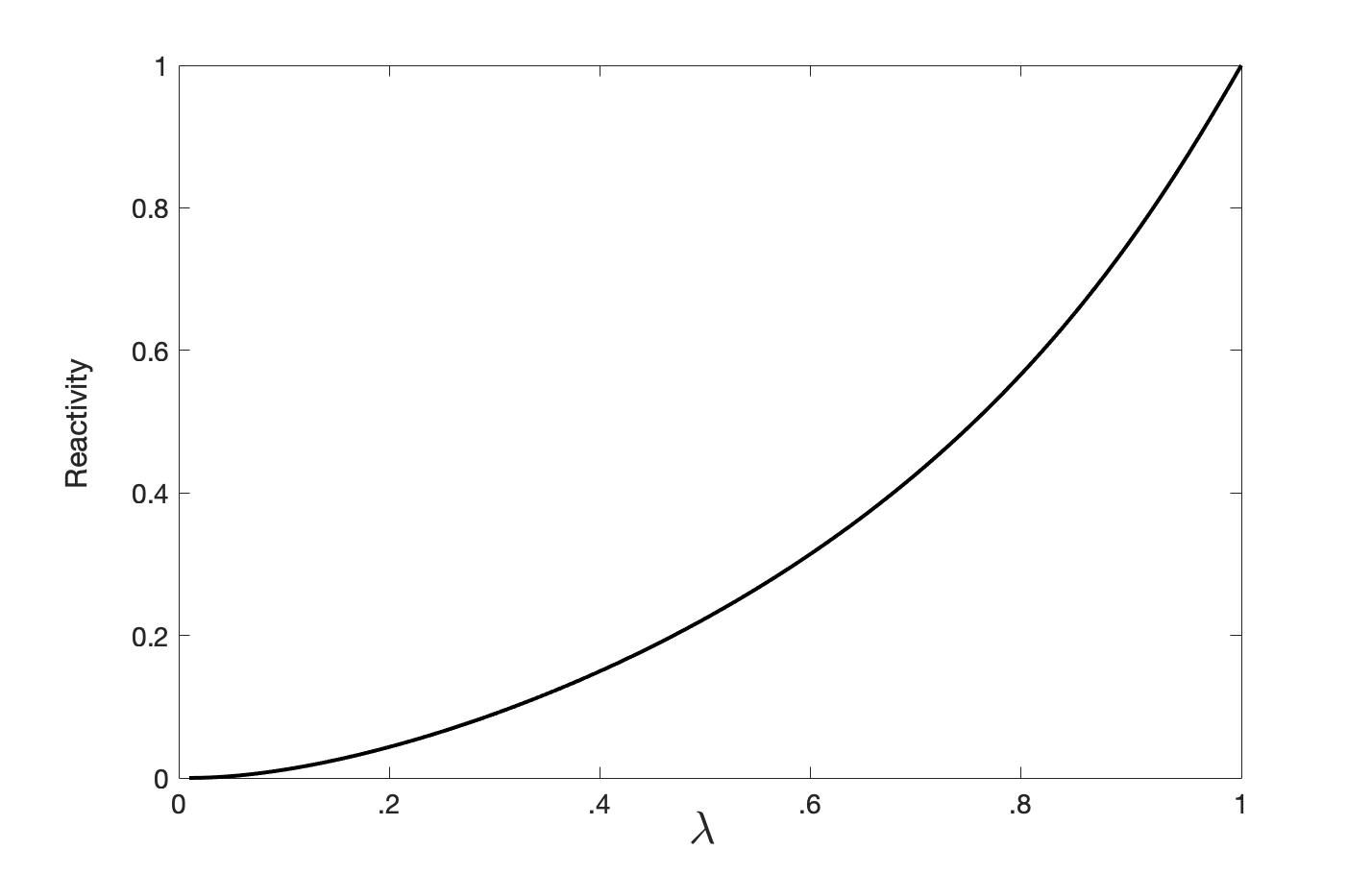}\
 \vspace{20pt}
 \caption{A plot of the reactivity (Eq.~\ref{eq:4reactivity}) for the 4--qubit Werner state (Eq.~\ref{Eq:WWWW}) as a function of the entanglement parameter $\lambda$. The reactivity was scaled with an affine transform so that ${\mathcal R}(\lambda=0)=0$ and ${\mathcal R}(\lambda=1)=1$   This example illustrates the scalability of the reactivity and that it yields a monotonically--increasing function in entanglement. }
 \label{fig:4werner}
 \end{figure}

We showed that an increase in the entanglement and correlation parameter $\lambda$ for the four--qubit Werner state yields correctly an increase in  its degree of correlation as a function of $\lambda$.  The reactivity in Eq.~\ref{eq:4reactivity} monotonically increases. This is illustrated in Fig.~\ref{fig:4werner}.  This is consistent with our results for the two--qubit example in Sec.~\ref{sec:werner}  where an increase in entanglement parameter gives a corresponding increase in curvature. 

\section{Conclusion}
\label{sec:fini}
We show in this paper that there is a suitable monotonic relationship between quantum correlation and reactivity.  This geometric definition is based on measurement outcomes and  gives us the opportunity to find a degree of quantum correlation emerging from its associated information geometry.  

We show for the two--qubit Werner state  that the reactivity qualitatively agrees with discord, although they are fundamentally different.  In Fig.~\ref{Fig:Werner} we compare concurrence and discord with our definition of reactivity for this mixed state.  While concurrence provides a measure for entanglement in that it is zero for separable states, it may be difficult to implement for higher–dimensional multipartite states. Both discord and reactivity are measures for quantum correlation and not entanglement.  Discord will always be an upper bound for reactivity; however, it may increase under LOCC in some cases \cite{Horedecki1}. Reactivity is non-increasing under LOCC \cite{Aslmarand:2019spie}.  

Reactivity is defined by a a weighted average of the surface--to--volume ratio over the space of measurements, and it appears to give us the opportunity to calculate a suitable degree of quantum correlation for higher-dimensional qudit states.  It is scalable to multipartite systems in the sense that it can be extended to a larger number of qubits.  As a measure of correlation, it has the advantage of being interpretation free unlike quantum discord for multipartite states.  Its is a relatively straightforward analytic function of the joint probability distribution.  Reactivity does not require any global minimization procedure or matrix inversion. In other words, it appears to us to be relatively easy to calculate in comparison to other measures of correlation. However,  the computational complexity of reactivity is dominated by  the computation of the  joint entropy over all the observers measurement outcomes and  scales exponentially in $d$ for a $d$--qubit quantum network. Nevertheless, this measure of quantum correlation appears to satisfy the properties required of such a measure, and perhaps it can be used as a mathematical tool to prove theorems in quantum computation? 

There are many questions remaining for us to explore, e.g.  
\begin{itemize}
\item Is it possible to calculate reactivity to high fidelity by using a random configurations of detectors for large multipartite systems? 
\item Can we couple our choice of measurements conditioned on previous measurements to probe quantum entanglement in quantum networks? 
\item Can we examine coarse-grained measurements of our qudit network, and its possible partitions?
\item Can we use a neural network or artificial intelligence algorithm to  probe optimal measurements to conduct on our quantum network? 
\item Is there a generalization of the Quantum Sanov's Theorem that will give exponential convergence in the fidelity of our measure of correlation?\cite{Vedral,Miller:2018}
\item  Does discord and reactivity capture the same aspects of quantum correlation?
\end{itemize}
Answering such questions would be important for quantifying the computational complexity of this method.

\section{Acknowledgments}

PMA and WAM would like thank support from the Air Force Office of Scientific Research (AFOSR).  WAM  research was supported under AFOSR/AOARD grant \#FA2386-17-1-4070. WAM wished to thank the Griffiss Institute and AFRL/RI for support under the Visiting Faculty Research Program.  Any opinions, findings, conclusions or recommendations expressed in this material are those of the author(s) and do not necessarily reflect the views of AFRL.


\begin{thebibliography}{999}
\bibitem{Preskill:2012}
 J. Preskill, Quantum computing and the entanglement frontier, Rapporteur talk at the 25th Solvay Conference on Physics ("The Theory of the Quantum World"), 19-22 October 2011; arXiv:1203.5813 [quant-ph].
 \bibitem{Bennett:1999}
 C. H. Bennett, D. P. DiVincenzo, C. A. Fuchs, T. Mor, E. Rains, P. W. Shor, J. A. Smolin, and W. K. Wootters, {\em Phys. Rev.} {\bf A 59}, 1070 (1999).
 \bibitem{Horodecki:2005}
 M. Horodecki, P. Horodecki, R. Horodecki, J. Oppenheim, A. Sen, U. Sen, and B. Synak-Radtke, {\em Phys. Rev.}  {\bf A 71}, 062307 (2005).
 \bibitem{Ali:2010}
 M. Ali, A. R. P. Rau, and G. Alber, {\em Phys. Rev.} {\bf A 81}, 042105 (2010); Erratum {\em Phys. Rev. } {\bf A 82}, 069902 (2010).
 \bibitem{Nisent:2006}
 J. Niset and N. J. Cerf, {\em Phys. Rev.}  {\bf A 74}, 052103 (2006).
 \bibitem{Lanyon:2008}
 B. P. Lanyon, M. Barbieri, M. P. Almeida, and A. G. White, {\em Phys. Rev. Lett.}  {\bf 101}, 200501 (2008).
\bibitem{Ekert}
 A. K. Ekert , ``Quantum cryptography based on Bell’s theorem,''  {\em Phys. Rev. Lett.} {\bf 67}, 661 (1991). 
 \bibitem{Charles H. Bennett}
C. H. Bennett and S. J. Wiesner, ``Communication via one- and two-particle operators on Einstein-Podolsky-Rosen states,''  {\em Phys. Rev. Lett.}  {\bf 69}, 2881 (1992).
\bibitem{Charles H Bennett}
C. H. Bennett, G. Brassard, C. Crépeau, R. Jozsa, A. Peres, and W. K. Wootters, ``Teleporting an unknown quantum state via dual classical and Einstein--Podolsky--Rosen channels,''  {\em Phys. Rev. Lett.}  {\bf70}, 1895 (1993).
\bibitem{Cleve}
R.  Cleve and H.  Buhrman, ``Substituting quantum entanglement for communication,''  {\em Phys. Rev.} {\bf  A  56}, 1201 (1997).
 \bibitem{Vedral}
V. Vedral, M. B. Plenio, K. Jacobs, and P. L. Knight, ``Statistical inference, distinguishability of quantum states, and quantum entanglement,'' {\em Phys. Rev.}  {\bf A56}, n.6 (1997) 4452. 
\bibitem{Horedecki1}
R. Horodecki, P. Horodecki,  M. Horodecki,  and K. Horodecki,  ``Quantum entanglement,'' {\em Mod. Phys.} {\bf81}, 865 (2009).
\bibitem{Wottward}
Scott Hill and William K. Wootters, ``Entanglement of a Pair of Quantum Bits,''
{\em Phys. Rev. Lett.} {\bf78}, 5022 (1997).
\bibitem{Zurek}
Harold Ollivier and Wojciech H. Zurek,  ``Quantum Discord: A Measure of the Quantumness of Correlations,''  {\em Phys. Rev. Lett.} {\bf88}, 017901 (2001).
\bibitem{Alber:2001} 
G. Alber, T. Beth, M. Horodecki, P. Horodecki, R. Horodecki, M. R\"otteler, H. Weinfurter, R. Werner, and A. Zeilinger, Quantum Information (Springer-Verlag, Berlin;  2001), ch. 5.
\bibitem{Rulli}
C. C. Rulli and M. S. Sarandy,  ``Global quantum discord in multipartite systems,''  {em Phys. Rev.} {bf  A 84}, 042109 (2011).
\bibitem{Schumacher}
B. W. Schumacher, ``Information and quantum nonseparability,''
{\em Phys. Rev.}  {\bf A 44}, 7047 (1991).
\bibitem{Miller:2018}
 W. A. Miller, ``Quantum Information Geometry in the Space of Measurements,''
  {\em Proc. SPIE} {\bf 10660} (2018) 16.
  \bibitem{Miller:2019Venn}
S. M. Aslmarand, W. A. Miller, V. S. Rana, and P. M. Alsing, ``Geometric measures of information for quantum state characterization," to be submitted (2019). 
\bibitem{Aslmarand:2019spie}
S.  M. Aslmarand, W. A. Miller \& V. S. Rana,  ``Properties of Quantum Reactivity for a Multipartite State,''  {\it Proc. SPIE} submitted (2019).
\bibitem{Wheeler:1990}
 J. A. Wheeler, ``Information, physics, quantum: the search for links,''  in Proc. 3rd Int. Symp. Foundations of Quantum Mechanics, Tokyo, 1989, Kobayashi, S., Ezawa, H., Murayama, Y., and Nomura, S., eds., 1898, 354–368, (World Scientific Publisher;  Phys. Soc. Japan; Tokyo ,1990).
\bibitem{Rolkin}
 Roklin, V. A., ``Lecture on the entropy theory of measure-preserving transformations,''  {\em Russian Mathematical Surveys}  \textbf{22}, 152 (1967).
\bibitem{Rajski}
 Rajski, C.,  ``A  metric space of discrete probability distributions,''  {\em Information and Control}  {\bf 4}, 373 (1961).
 \bibitem{Miller}
Caves, C., Kheyfets, A., Lloyd, S., Miller, W. A., Schumacher, B., and Wootters, W. K., Notes." unpublished (April 18, 1990).
 \bibitem
{Zurek:1989} W. H. Zurek, “Thermodynamic cost of computation, algorithmic complexity and the information metric,” {\em Nature} {\bf 341}, 119 (1989).
\bibitem{Shannon}
CE Shannon,  A mathematical theory of communication ,Bell system technical journal, 1948 - Wiley Online Library.
\bibitem{Thiago}
Thiago R. de Oliveira , ``Quantum Correlations in Multipartite Quantum Systems,''  in Lectures on General Quantum Correlations and their Applications, eds. Fanchini, F. F, Soares P., de~Oliveira D., Adesso, G. (Springer International Pubs.,  2017) ch. 5, pp. 87--103; arXiv:1706.03101v1, 9 Jun 2017.
\bibitem{Werner}
Reinhard F. Werner, ``Quantum states with Einstein-Podolsky-Rosen correlations admitting a hidden-variable model,''  {\em Phys. Rev.}  {\bf A 40}, 4277 (1989).

\end{thebibliography}
\end{document}